\documentclass{PoS}
\usepackage{wasysym}
\usepackage{booktabs}

\usepackage[switch]{lineno}

\title{Construction and test of high precision drift-tube (sMDT) chambers for the ATLAS muon spectrometer}

\ShortTitle{Construction and test of high precision drift-tube (sMDT) chambers for the ATLAS muon spectrometer}

\author{Sebastian NOWAK\\
  Max-Planck-Institut f\"ur Physik\\
  E-mail: \email{nowak@mppmu.mpg.de}}
\author{Oliver KORTNER\\
  Max-Planck-Institut f\"ur Physik\\
       E-mail: \email{kortner@mppmu.mpg.de}}
\author{Hubert KROHA\\
  Max-Planck-Institut f\"ur Physik\\
  E-mail: \email{kroha@mppmu.mpg.de}}
\author{Philipp SCHWEGLER\\
  Max-Planck-Institut f\"ur Physik\\
  E-mail: \email{philipp.schwegler@cern.ch}}
\author{\speaker{Federico SFORZA}%
  Max-Planck-Institut f\"ur Physik\\
  E-mail: \email{federico.sforza@cern.ch}}

\abstract{For the upgrade of the ATLAS muon spectrometer in March 2014 new muon tracking chambers (sMDT) with drift-tubes of 15 mm
diameter, half of the value of the standard ATLAS Monitored Drift-Tubes (MDT) chambers, and 10~$\mu$m positioning accuracy of the sense wires have
been constructed. The new chambers are designed to be fully compatible with the present ATLAS services but, with respect to the previously
installed ATLAS MDT chambers, they are assembled in a more compact geometry and they deploy two additional tube layers that provide redundant
rack information. The chambers are composed of 8 layers of in total 624 aluminium drift-tubes. The assembly of a chamber is completed within a
week. A semi-automatized production line is used for the assembly of the drift-tubes prior to the chamber assembly. The production procedures and
the quality control tests of the single components and of the complete chambers will be discussed. The wire position in the completed chambers
have been measured by using a coordinate measuring machine.}

\FullConference{Technology and Instrumentation in Particle Physics 2014\\
		 2-6 June, 2014\\
		 Amsterdam, the Netherlands}

\begin{document}
%\linenumbers
\section{Introduction}

This paper describes the production, the quality assurance and the geometry characterization of two new muon tracking chambers constructed at the Max-Planck-Institut f\"ur Physik (MPI, M\"unich) and designed for the upgrade of the ATLAS Muon Spectrometer (MS) during the 2013-2014 shutdown of the Large Hadron Collider (LHC) at CERN. 

The ATLAS MS is immersed in a toroidal magnetic field and it is instrumented almost completely with three stations (Inner, Medium, and Outer) of Monitored Drift-Tube (MDT) chambers for precision tracking in the bending plane of the traversing charged particles. The new chambers are designed to improve the 3-station coverage of the MS in the so-called BME region where two elevator shafts, used for the access to the inner part of the ATLAS detector, limit the MDT tracking stations to only two. The momentum resolution of simulated 1~TeV muons passing trough the BME region would be improved by about 50\% thanks to the additional chambers. 

\begin{figure}[!ht]
  \begin{center}
    \includegraphics[width=0.65\textwidth]{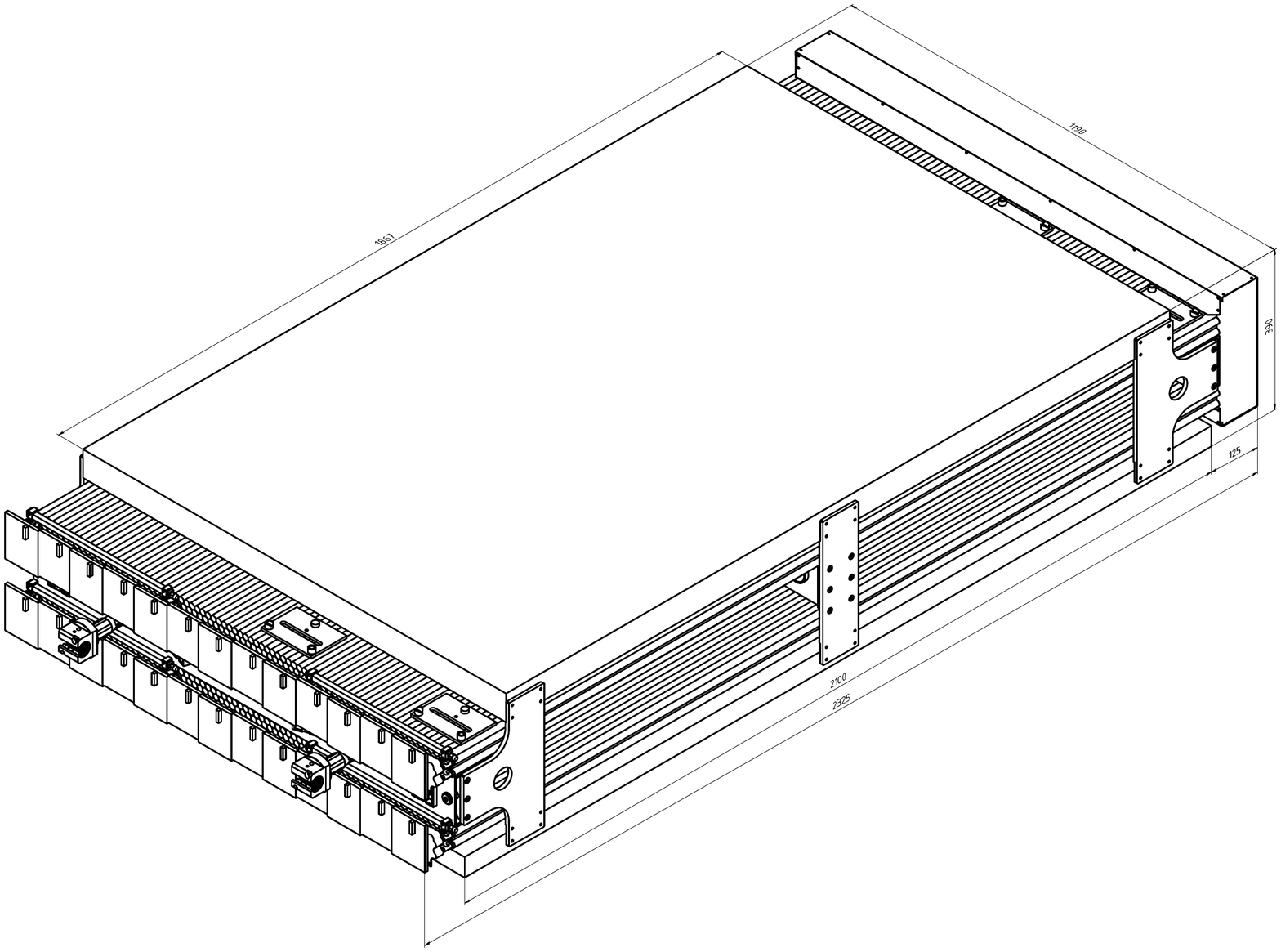}
    \caption{Schematic view of a new Monitored Drift Tube (MDT) chamber designed for upgrade of the ATLAS MS during the 2013-2014 LHC shutdown. The chamber uses 624 drift-tubes of 15 mm diameter arranged in 2 multi-layers sustained by an aluminum spacer frame. Each multi-layer is composed by 4 layers of 78 drift-tubes.}\label{fig:chamber}
  \end{center}
\end{figure}

Figure~\ref{fig:chamber} shows a schematic view of the chambers designed for the BME upgrade. With respect to the MDT chambers currently instrumenting the ATLAS MS, the newly produced chambers (also named sMDT) are characterized by a technology using ``small'' drift-tubes, with diameter one half of the standard 30~mm diameter drift-tubes used for the ATLAS MDT chambers. This allows a factor 10 higher tolerance to $\gamma$ and neutron backgrounds~\cite{sMDT_bkg}, redundant track information, and a compact design suitable for the upgrade of the hard-to-access MS regions. At the same time the sMDT chambers, operated with a voltage of 2730 V on the anode wire and with a gas mixture Ar:CO$_2$ (93:7) at 3 bar pressure, are fully compatible with the present services and electronics.

The chambers themselves are build by 624 aluminum, 2.2~m long,  drift-tubes arranged in two multi-layers each composed by 4 layers of 78 tubes, for a total width of 1.2~m. 
The quality assurance and the precision assembly of so many individual components are two of the main challenges of the sMDT production. The two multi-layers are sustained by an aluminum spacer frame equipped with four RasNiK~\cite{rasnik} optical alignment lines to monitor deformations with a few micron precision.  
Rail supports are mounted on the frame to allow the sliding of the chamber from the parking position, that still allows the access to the elevator shaft area, to the data-taking position. During data taking, precision alignment platforms mounted on the chamber allow a nominal positioning accuracy of 30 $\mu$m with respect to the rest of the ATLAS MS.

\section{Tube Production and Quality Assurance}

Each drift-tube is a multi-component device~\cite{sMDT_construction} that must be assembled and tested for tight quality criteria. In particular, each tube must satisfy to three important requirements: correct crimping and tensioning  (with tension of 3.5$\pm$0.15~N) of the anode wire to ensure uniformity and centrality of the wires in the aluminum tubes, low leakage current ($<1$~nA) to ensure noise rates below threshold, and gas pressure tightness (leak rate $0.4<10^{-8}$~bar$\cdot$L$/s$) to avoid changes in the drift properties of the gas.

\begin{figure}[!ht]
  \begin{center}
    \includegraphics[width=0.49\textwidth]{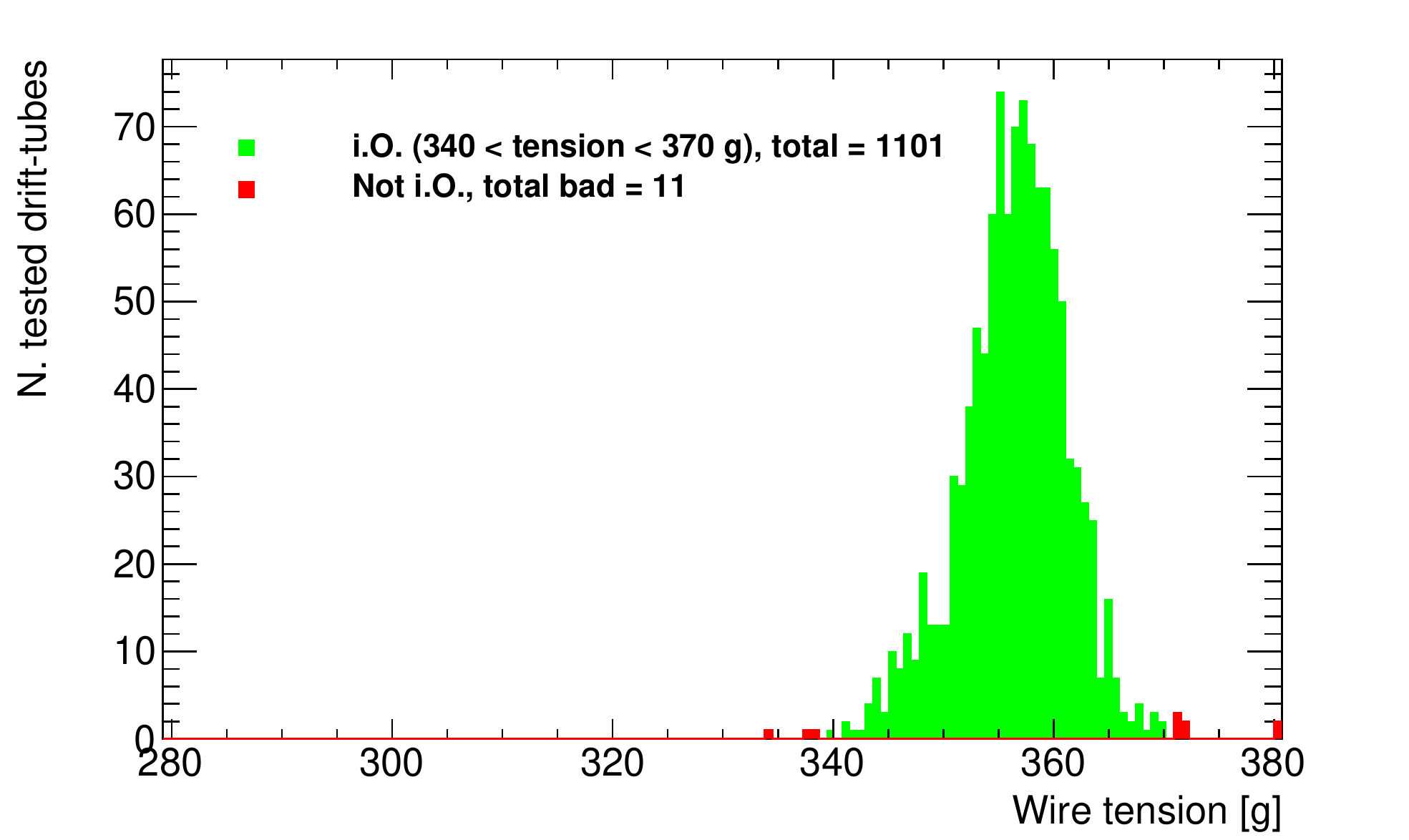}\\
    \includegraphics[width=0.49\textwidth]{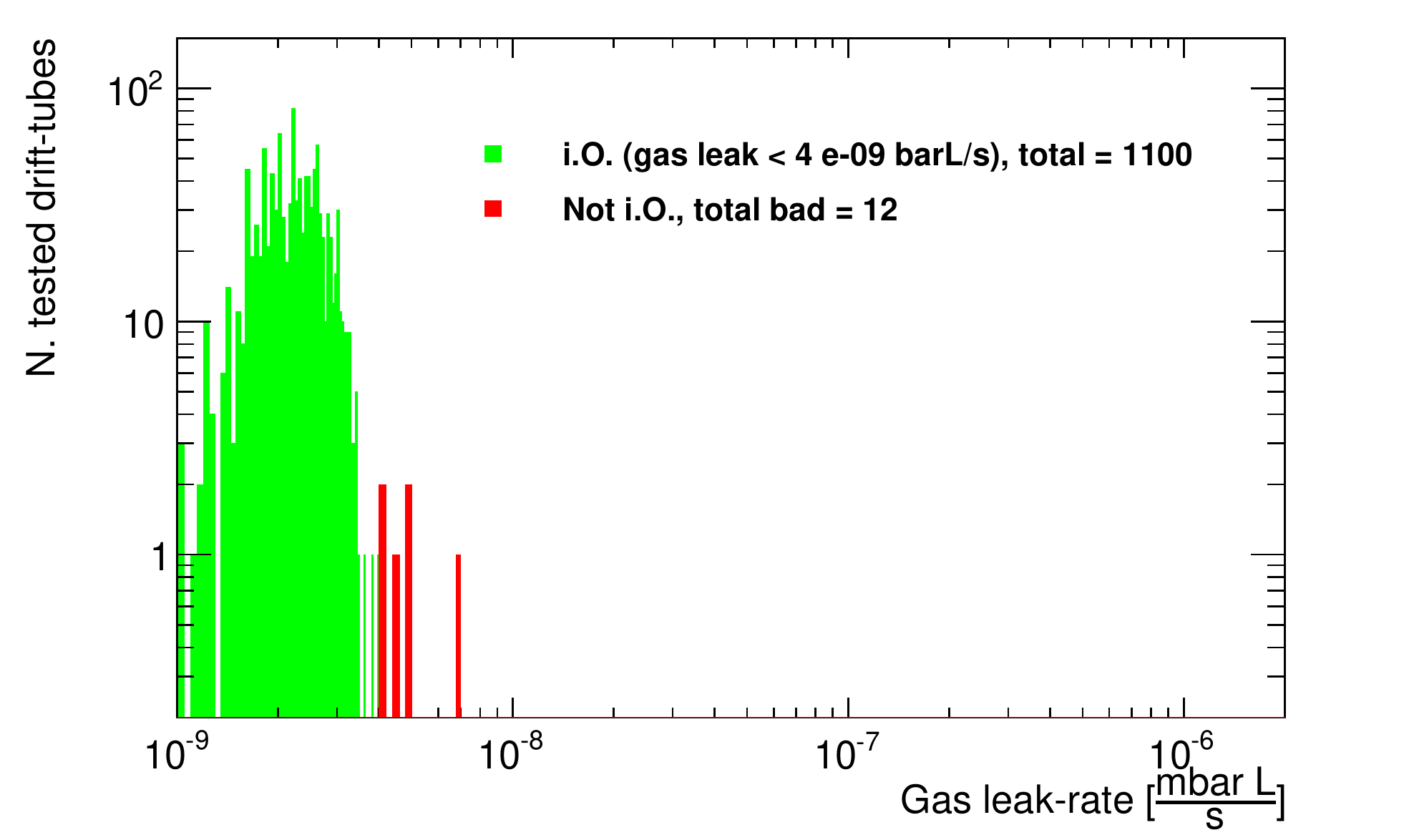}
    \includegraphics[width=0.49\textwidth]{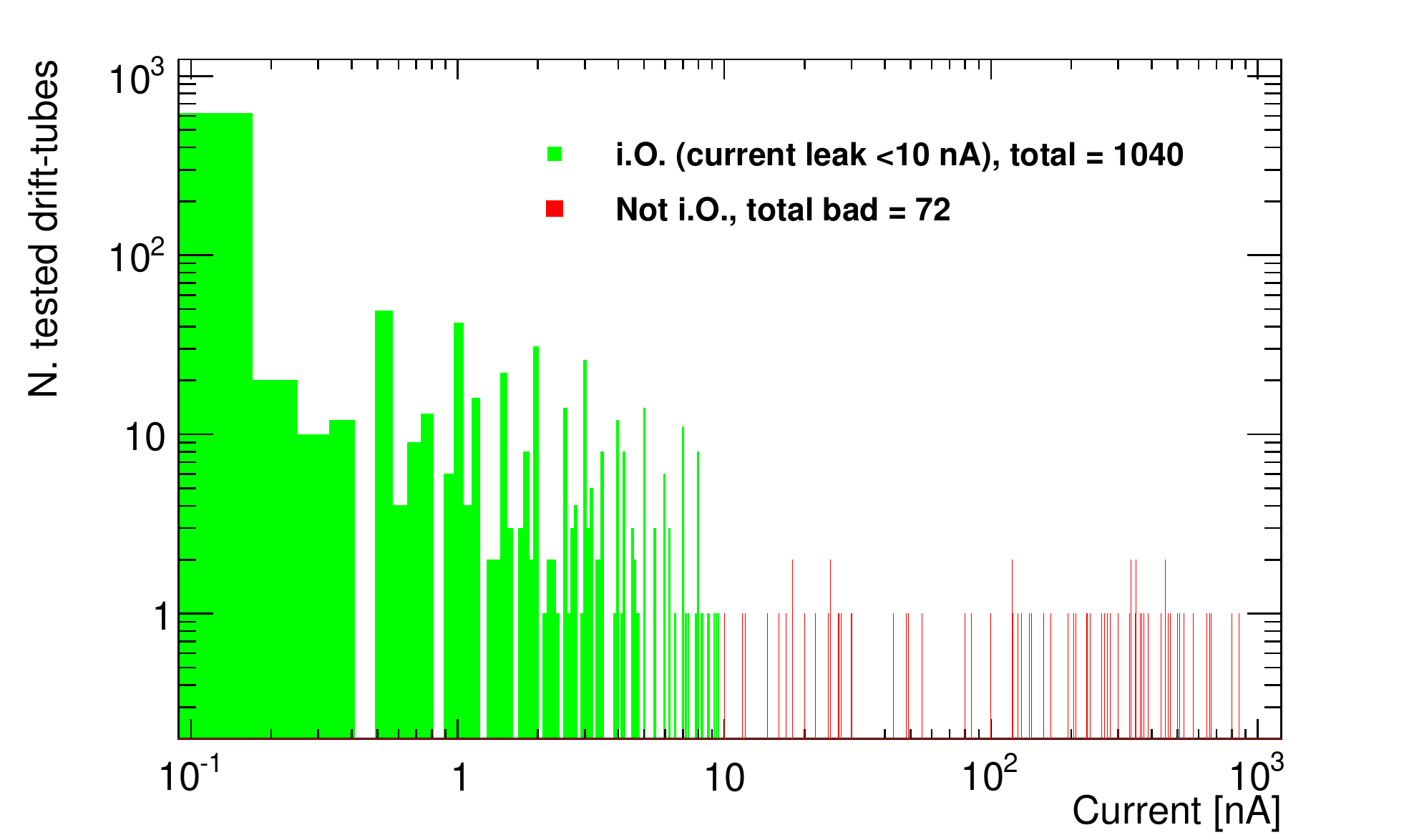}
    \caption{Result of the quality assurance checks performed on the last 1112 drift-tubes produced for the construction of. The upper plot shows the anode wire tension force, the optimal value of 3.5$\pm$0.15 N has been increased to account for 0.05 to 0.07 N relaxation of the wire observed in the days after the crimping. The lower left plot shows gas leak rate of the tested tube when injected with 3 bar pressure of Ar:CO$_2$ (93:7) gas. The bottom right plot shows the measured leakage current for the drift tube under pressure and for an anode    voltage set to 3015 V.}\label{fig:QA}
  \end{center}
\end{figure}

The drift-tube assembly and checks are performed in clean environment using a production line composed by two semi-automated wiring and testing stations. The values resulting from the tests, associated to the bar-code identifier of each drift-tube, are stored in an MySQL database. Figure~\ref{fig:QA} shows the results of the quality assurance checks on the last 1112 drift-tubes produced. Each tube not passing the requirements is discarded. The leak current test failure rate of approximately $6.5$\% can be improved by more strict requirements on the handling and storing of the not tested drift-tubes.

At regime and operated by three persons, the current setup has a production and test capability of 50$-$60 drift-tubes per day.

\section{High Precision Chamber Assembly}

The ATLAS experiment requires the wire positions to be known with an uncertainty of 20~$\mu$m to ensure optimal tracking resolution performance of the chamber. To achieve such precision  goal two jigging, each composed by five metal blocks with pin supports and precision surfaces, are mounted on a very stable granite table and are used to sustain the drift-tubes on the two sides  (named North side, and South side) and glue them in position. The precision surfaces have been machined with an accuracy of 15~$\mu$m and the pin supports ensure the positioning of the wires as these are tensed at the center of a micron-precise cylindrical brass end-plug. 

The complete assembly of a sMDT chamber is performed in five working days with the help of a programmable gluing machine: on day 1 the first two layers of 78  drift-tubes are positioned and glued; on day 2 the first multi-layer is completed with other two drift-tube layers and from the jigging; on day 3 the first two layers of the second multi-layer are positioned and glued; on day 4 the second multilayer is completed and a dry-test of the frame support positioning is performed to check the optical alignment lines mounted on the frame; on day 5 the frame support is glued to the second multilayer that is still in position and the first multilayer is glued on the top.

The assembly procedure is critical because no repairs are possible after the gluing of the multi-layers. To achieve the maximal precision and reliability the metal plates of the jigging are inspected before every step and the parameters of each drift-tube are checked in the database before positioning.
The successive steps in the chamber construction are the gas distribution and electronic system montage. These phases are less critical because repair and substitution of defective components can be done also at later stage, however the procedure may be time consuming because of the high number of components to be connected and tested.

\section{Evaluation of Chamber Geometry and Precision}\label{sec:geo}

After the chamber assembly but before the gas distribution system montage, the geometry of the chamber is tested with a 3D Coordinate Measuring Machine (CMM) used to measure the wire positions, through the brass end-plugs, with a precision below 5$\mu$m. 

The wire position measurements are used to characterize the chamber geometry with a wire-grid for the North and for the South sides.  Five parameters\footnote{According to the ATLAS reference system the $x$ coordinate is defined in the direction of the drift-tube wires, the $y$ coordinate is vertical and growing with the drift-tube layer count, and  the $z$ coordinate is  along the single drift-tube layer.} describe the wire-grid for each side: two wire pitches ($z-$pitch and $y-$pitch along $y$ and $z$), the $z$ displacement of even layers with respect to the odd ones ($z$ odd/even), the $z$ and $y$ distance between the two multi-layers ($z$ ML and $y$ ML). These are extracted from the CMM measurements with a minimum $\chi^2$ fit, then the single  $z$ and $y$ positions of the wires are compared to the fitted grid values and the standard-deviations, $\sigma_z$ and  $\sigma_y$, of an iterative Gaussian fit of the distribution is used to evaluate the geometrical precision of chamber. Table~\ref{tab:geometry} shows the geometrical characterization of the two sMDT chambers BME-A and BME-C produced for the BME upgrade (they are named according to the ATLAS MS installation side). 

\renewcommand{\tabcolsep}{1.3pt}

\begin{table}[h!]
  \centering
{\footnotesize
  \begin{tabular}{ccccccc}\toprule
    par. [mm] & BME-A N. &BME-C N.& Nominal N.  & BME-A S. &BME-C S.& Nominal S.  \\\midrule
     $z-$pitch  & 15.09950$\pm (1\cdot 10^{-5})$ & 15.09930$\pm (1\cdot 10^{-5})$ &  15.100 & 15.09900$\pm (1\cdot 10^{-5})$ & 15.09880$\pm (1\cdot 10^{-5})$ &15.100\\ 
     $y-$pitch  & 13.0956$\pm (1\cdot 10^{-4})$ & 13.0972$\pm (1\cdot 10^{-4})$ & 13.095 &  13.0857$\pm (1\cdot 10^{-4})$ & 13.0851$\pm (1\cdot 10^{-4})$&  13.085 \\
     $z$ odd/even &  7.5529$\pm (3\cdot 10^{-4})$ &  7.5511$\pm (3\cdot 10^{-4})$ & 7.550      & 7.5509$\pm (3\cdot 10^{-4})$ & 7.5461$\pm (3\cdot 10^{-4})$& 7.550\\
     $z$ ML & 0.0089$\pm (3\cdot 10^{-4})$ & -0.0073$\pm (3\cdot 10^{-4})$ & 0 & 0.0283$\pm (3\cdot 10^{-4})$&   0.0095$\pm (3\cdot 10^{-4})$& 0 \\
     $y$ ML &  135.3417$\pm (5\cdot 10^{-4})$  & 135.3562$\pm (5\cdot 10^{-4})$ &135.347 & 135.2815$\pm (5\cdot 10^{-4})$ &  135.2996$\pm (5\cdot 10^{-4})$  &135.271\\
     $\sigma_z$& $6\cdot 10^{-3}$ & $6\cdot 10^{-3}$ &$20\cdot 10^{-3}$ &  $7\cdot 10^{-3}$&  $7\cdot 10^{-3}$& $20\cdot 10^{-3}$\\
     $\sigma_y$& $12\cdot 10^{-3}$ & $13\cdot 10^{-3}$ &$20\cdot 10^{-3}$ &  $8\cdot 10^{-3}$&  $8\cdot 10^{-3}$& $20\cdot 10^{-3}$\\
    \bottomrule
  \end{tabular}
}
  \caption{Parameters describing the wire-grid of the North (N.) and South (S.) side of the constructed sMDT chambers. Parameters are  derived from a minimum $\chi^2$ fit of the wire position measurements obtained with a 3D CMM. The $\sigma_z$ and $\sigma_y$ parameters, used to evaluate the precision of the chamber geometry, are extracted  from an iterative Gaussian fit of the distribution of the residuals of the $z$ and $y$ wire coordinates with respect to the wire-grid.}
  \label{tab:geometry}
\end{table}

Using the described geometry parametrization of the chambers we obtain a construction precision well below the stringent requirement of 20$~\mu$m. The two newly constructed sMDT chambers result as the most precise produced chambers of this size.

\section{Conclusions}
Two precision MDT chambers using small (15 mm $\diameter$) drift-tube technology have been developed and produced at MPI, in Munich, for the upgrade of the ATLAS MS during the 2013-2014 shutdown of the LHC at CERN. The sMDT chamber production line consists of drift-tube production and quality testing stations with a capacity of 50 drift-tube a day, and an area for the precision assembly and gluing of the chambers. The current setup of the production line can be used for future upgrades of the ATLAS MS using the sMDT technology. The geometry of the newly produced chambers has been evaluated with 3D coordinate measuring machine revealing a construction precision of approximately $10$~$\mu$m resulting in the most precise MDT chambers of these dimension ever produced.

\end{document}